\documentclass[authoryear,12pt,3p,letter]{jowarticle}
\usepackage{graphicx}
\usepackage{amsthm,amsmath}
\usepackage{amssymb}
\usepackage{amsfonts}
\usepackage{natbib}
\usepackage{setspace}
\usepackage[all]{xy}
\usepackage{enumitem}
\usepackage{titlesec}
\usepackage{mathrsfs}
\makeatletter
\renewcommand\section{\@startsection{section}{1}{\z@}{-3.25ex plus -1ex minus -.2ex}{1.5ex plus .2ex}{\normalsize\bf}}
\renewcommand\subsection{\@startsection{subsection}{2}{\z@}{-3.25ex plus -1ex minus -.2ex}{1.5ex plus .2ex}{\normalsize\bf}}
\renewcommand\subsubsection{\@startsection{subsubsection}{3}{\z@}{-3.25ex plus -1ex minus -.2ex}{1.5ex plus .2ex}{\normalsize\bf}}
\makeatother

\newtheorem{innercustomgeneric}{\customgenericname}
\providecommand{\customgenericname}{}
\newcommand{\newcustomtheorem}[2]{%
  \newenvironment{#1}[1]
  {%
   \renewcommand\customgenericname{#2}%
   \renewcommand\theinnercustomgeneric{##1}%
   \innercustomgeneric
  }
  {\endinnercustomgeneric}
}

\newcustomtheorem{prop*}{Proposition}
\newcustomtheorem{cor*}{Corollary}

\begin{document}
\begin{frontmatter}
\title{Two Dogmas of Dynamicism}
\author{James Owen Weatherall}\ead{weatherj@uci.edu}
\address{Department of Logic and Philosophy of Science\\ University of California, Irvine}
\begin{abstract}
I critically discuss two dogmas of the ``dynamical approach'' to spacetime in general relativity, as advanced by Harvey Brown [\emph{Physical Relativity} (2005) Oxford:Oxford University Press] and collaborators.  The first dogma is that positing a ``spacetime geometry'' has no implications for the behavior of matter.  The second dogma is that postulating the ``Strong Equivalence Principle'' suffices to ensure that matter is ``adapted'' to spacetime geometry.  I conclude by discussing ``spacetime functionalism''.  The discussion is presented in reaction to and sympathy with recent work by James Read [``Explanation, geometry, and conspiracy in relativity theory'' (20??) \emph{Thinking about Spacetime} Boston: Birk\"auser].
\end{abstract}

\end{frontmatter}
\doublespacing

\section{Introduction}

In a recent paper, James \citet{ReadIreny} has proposed a detente in an active and ongoing debate in the foundations of spacetime theories, between defenders of the ``geometrical approach'' to spacetime structure and those of the ``dynamical approach'', introduced by \citet{Brown+Pooley1, Brown+Pooley2} and elaborated and defended by \citet{Brown}.\footnote{For a recent, but somewhat one-sided, review of this literature, see \citep{Brown+Read}. The dynamical view is clearly articulated in the papers cited in the main text.  But since the ``geometrical view'' is the received view, it is somewhat difficult to identify a \emph{locus classicus} for its statement. I would argue that the view goes back at least to \citet{Weyl}. It is also the view implicitly found in, for instance, \citet{Hawking+Ellis}, \citet{Wald}, and \citet{MalamentGR}; see also \citet{SteinGrunbaum}, discussed in the final section of this paper.  \citet{Friedman}, \citet{Torretti}, and \citet{Maudlin} are all sometimes cited as defenders of the geometrical view.}  The geometrical view, Read suggests, has been given an unfair treatment by recent defenders of the dynamical view---including Read and his collaborators \citep[e.g.][]{Miracles, Misconceptions, Brown+Read}.  Really, there are two versions of the geometrical view that Read detects in the literature: what he calls the ``unqualified'' and ``qualified'' geometrical approaches.  (I will describe these below.) The first of these, he argues, is untenable.  But the second, which he tentatively attributes to \citet{Maudlin} and more confidently attributes to me \citep{WeatherallConservation}, is tenable after all, he says.  In fact, Read thinks, on reflection it is hard to see what the disagreement between the qualified geometrical view and the dynamical view is supposed to be, at least in the context of general relativity.\footnote{In the context of special relativity, conceived as a separate theory from general relativity, he thinks that both views are tenable, but that they are clearly distinct and he prefers the dynamical view.  See \citep{ReadConstructive} for further discussion of issues that are salient to the debate in special relativity, but which will not concern me in the present article.}  Indeed, Read argues that my \citeyear{WeatherallConservation} is best read as endorsing not only the dynamical approach, but also a version of spacetime functionalism, as introduced by \citet{KnoxEffective,KnoxFunctionalism}, which Read sees as an ``extension'' of the dynamical approach according to which spacetime is whatever plays the spacetime role, and ``the spacetime role is played by whatever defines a structure of local inertial frames'' \citep[p.22]{KnoxFunctionalism}.\footnote{See \citet{Baker} and \citet{Read+Menon} for critiques of Knox's view, and \citep[fn. 40]{WeatherallConservation} for a discussion how I see the relationship between my views as discussed there and Knox's.}

There is much to admire in Read's discussion and I think his ireny is a sign of considerable progress all around.  To the extent that I have summarized it so far, I largely agree with what he writes, and in what follows, I will offer an analysis that is very similar to his.  But I will come at it from a different perspective---one closer, I think, to the received (that is, geometrical) view, though I am reluctant to take on such a mantle because I agree with Read that the two sides, in their most careful moments, are not so clearly different on the issues that are generally taken to define the debate.  I think there is some value in rehashing these issues from this other perspective, both because it can aid in translating between somewhat different ways of speaking about the foundations of general relativity; and because I think it will help isolate where disagreements persist.

The paper will proceed as follows.  I will begin by restating some key points from Read's paper in a different way; as I go, I will explain how I take the issues as I raise them to relate to what Read writes.  I will then discuss one important point on which I think there is still disagreement.  As will be clear, setting aside quibbles about terminology, this disagreement has little to do with the headline questions that have marked the debate, such as whether geometry is ``reducible'' to dynamics or whether geometry ``explains'' the behavior of matter, and much more to do with the details of what the relationship between dynamics and geometry is supposed to be.\footnote{One might wonder whether the remaining disagreement I identify is really so far from those headline issues---or if rather, what I am attempting here is to isolate what has always been the core disagreement, from which the other debates have followed.  Perhaps. But if so, I think it is doubly important to focus on this particular disagreement, because I think it is one that can be substantially resolved with new technical work.  Thank you to Sam Fletcher for raising this point.} Along the way, I will offer critical comments on two ``dogmas'' of the dynamical approach: one, already questioned by Read in somewhat different terms, is the idea that positing a certain geometrical structure for spacetime should be understood to have no consequences for the dynamics of matter; and the second, which I think has received too little attention in the literature, is the idea that the ``strong equivalence principle'', as stated by Brown, Read, Knox, and others, suffices to explicate the intended relationship between matter and geometry in general relativity.  I will conclude with some remarks concerning the bearing of these arguments on spacetime functionalism as discussed by Knox and others.

\section{Isn't It Irenic?}

Consider the following characterization of general relativity:\footnote{I take for granted the technical details of general relativity in what follows; for background using similar notation, see \citet{MalamentGR} or \citet{Wald}.}
\begin{quote}General relativity is a theory of spacetime geometry according to which:\\
{\bf [(g)eometry]} Spacetime has the structure of a Lorentzian 4-manifold, $(M,g_{ab})$, the curvature of which is related to the distribution of stress-energy in the universe by Einstein's equation; and \\
{\bf [(d)ynamics]} the dynamical evolution of matter is adapted to this geometry.\end{quote}
This characterization of general relativity is stronger than one sometimes seen in the literature (for instance, it is stronger than what \citet[\S 2.3.2]{ReadIreny} provides), since it includes (i) an explicit interpretation of the spacetime metric, namely as a description of physical geometry; and (ii) an analysis of what it means to give the metric this interpretation, viz. (d).\footnote{For \citet{ReadIreny}, general relativity, or a ``general relativistic theory'', is a theory in which there are matter fields on Lorentzian manifolds, where the matter fields and metric are required to satisfy some system of differential equations that includes Einstein's equation sourced by some stress-energy tensor associated with the matter fields.  This statement is similar to mine, except that no constraints are made, in the first instance, on what that system of equations can be, except that it include Einstein's equation.  Note that both descriptions of the theory are vague on some important issues, such as how a stress-energy tensor is to be associated with matter.}  But I also think it is a fair characterization of what one might (or should) mean by a ``geometrical'' view of general relativity.

Now, given this gloss on the theory, one may pose the following two questions.
\begin{enumerate}[start=1,label={(\Roman*)}]
\item Is (d) independent of (g)?
\item Can (d) be made precise, in the sense of giving conditions on matter dynamics such that those dynamics are ``adapted'' to the spacetime geometry?
\end{enumerate}

Taking (I) first: this question is the central issue at stake for the dynamical view.  In particular, \citet{Brown} argues (in different terms) that (d) \emph{is} independent of (g); and moreover, because (d) is independent of (g), (g) alone cannot ``explain'' how the spacetime metric, $g_{ab}$, comes to have its ``chrono-geometric significance''---that is, (g) alone does not imply anything about the behavior of light rays, small bodies, or stylized measuring devices (``rods and clocks'').\footnote{Here and in what follows, I am taking certain stylized facts about the behavior of matter---for instance, that small massive bodies follow timelike geodesics, light rays follow null geodesics, etc.---to be necessary for matter to be ``adapted'' to geometry.  (I will have much more to say on this issue below.)  To connect my way of putting things in the main text to other discussions in the literature, I take \citet{Brown} to be claiming that (d) is independent of (g) when he argues that it is only if one assumes the ``strong equivalence principle'', discussed below, that one can explain these behaviors; and likewise, when \citet{Miracles} use the ``two miracles'', also discussed below, as a litmus test for distinguishing the dynamical view from the geometric view.  On this last point, what makes the ``two miracles'' miraculous is precisely that they do not follow from any other principles of general relativity---or more specifically, from (g)---and must be taken as brute facts.}   Conversely, Brown and collaborators understand the advocate of the ``geometrical view'' to \emph{deny} that (d) is independent of (g), by maintaining that (d) follows from (g), or at least, that (g) ``explains'' (d).  Read cites authors such as Michael \citet{Friedman} and Tim \citet{Maudlin} who speak of spacetime geometry explaining, for instance, why light rays follow null geodesics in curved spacetime as committed to this position; in earlier work, Brown and Pooley use Roberto \citet{Torretti} as a foil.

Put in these terms, whether the defender of the dynamical view is correct that (d) is independent of (g) depends on what one means by (g).  If by (g) one means---as, apparently, the defenders of the dynamical approach do, at least when they criticize the geometrical approach---some bare metaphysical hypothesis about what the ``true'' geometry of spacetime is, independent of the behavior of any physical system in that spacetime,\footnote{That is: any \emph{other} physical system aside from the fields associated with geometry, since one might take the field $g_{ab}$ as a physical system.  Thanks to an anonymous referee for encouraging this clarification.} then it seems true that (d) does not follow from (g).\footnote{This interpretation of (g) is not completely contrived: for instance, it is closely related to the understanding of ``spacetime geometry'' at issue when some philosophers and physicists entertain the possibility that spacetime truly has a Galilean structure even though matter behaves ``as if'' spacetime is Minkowskian.  In such cases, one makes a claim about the ``true'' spacetime geometry and leaves open what bearing that has for dynamics.  See the discussion at the end of this section of what Read calls the ``unqualified geometrical approach''.}   Indeed, on this interpretation, (g) is explicitly understood in a way that denies any link with (d).  But I want to raise the possibility that this is not the only thing that one \emph{might} mean by postulating a ``spacetime geometry'' as in (g).  Another meaning that one might give the expression ``spacetime geometry'' is precisely that it is whatever geometry the dynamical evolution of matter is adapted to, in a sense that is at least roughly understood from examples such as Maxwell's theory in curved spacetime.  From this perspective, (d) is simply a restatement, or maybe an explication, of (g); and (d) \emph{would} follow from (g), properly understood.

How might we understand (g) so that (d) follows from it?  The key is to recognize the often subtle ways in which our mathematical terms are chosen to reflect intended interpretations. We speak of the ``length'' of a vector and the ``length'' of a table; the second use of ``length'' refers to a property of a physical object, whereas the first refers to a property of a mathematical one, where one borrows the term because it invokes the idea that the length of a vector can represent physical lengths.  But whatever the original motivation for choosing mathematical language may be, the meaning of such terms may shift as applications do---and to understand them, we must attend carefully to the context.\footnote{\citet{SteinGrunbaum}, discussed in the final section of this paper, offers a more detailed historical analysis of how geometrical language came to have its present significance in relativity theory.  \citet{Wilson} provides a more general discussion of how meaning shifts with new applications, particularly in the context of (applied) mathematics.} To say spacetime has a certain geometrical structure according to general relativity is, in the first instance, to assert that there are facts concerning space and time that are represented, mathematically, by a structure we associate with ``geometry''---viz. a pseudo-Lorentzian metric.  These facts are given names such as ``length'', ``angle'', ``duration'', ``distance'', etc. because those are the words we use in the mathematical context for the corresponding quantities determined by a metric (borrowing, of course, from prior physical usage).  We can summarize all of this by saying (g) asserts that there are facts about length, angles, etc. that obtain between events in space and time.\footnote{As I hope is clear in the main text, including the paragraph that follows, I am not advocating for, or attributing to the ``geometricist'', any particular view about, say, spacetime substantivalism; the significance I attribute to claims about ``length'', ``angle'', etc. are compatible with a broad range of positions in that debate.  (For more on my own view in that connection, see \citep{WeatherallHoleArg,WeatherallStein}.) On a related note, \citet{Acuna} describes views along the lines of the one I describe in the text as ``absolutist''.  He may use terms as he likes, but it is worth emphasizing that this use of ``absolutist'' is importantly different from its usage in connection with, say, Newton's views on absolute space and time. (See also \citet{ReadConstructive} for a discussion of Acu\~na in this regard.)}

But what is the physical significance of such assertions in the present context?  To answer that, we observe that the properties of matter represented by matter fields in general relativity evolve according to certain equations that \emph{also} invoke precisely those mathematical quantities that we call ``lengths'', ``angles'', etc., as determined by some metric.  For instance, the wave equation may be understood as asserting that ``the integrated length of the gradient of field strength is extremal in any compact region.''\footnote{Here I am invoking the fact that the wave equation is the Euler-Lagrange equation extremizing a certain action.}  Likewise, the source-free Maxwell equations say ``the integrated magnitude of field strength is extremal in any compact region'', where ``magnitude'' here is an analogue of length applying to tensors at a point.  (Or consider an older example: ``the acceleration of a body under the gravitational influence of another body is proportional to the square of the distance between them.'')  Asserting that space and time have a certain geometry, then, is just to assert a link between these various appearances of ``lengths'', ``angles'', etc. in different parts of the theory: that is, it is to say that the geometrical relations that matter dynamics advert to are precisely the geometrical relations one has postulated of the world and summarized with the metric.\footnote{There are other reasons, too, to think that ``spacetime geometry'' should be interpreted in this stronger way. \citet{Acuna} and \citet{Myrvold}, for instance, argue that the link between geometry and dynamics is analytic, and thus that all one \emph{could} mean by attributing a certain geometry to space and time is to say that matter behaves in particular ways adapted to that geometry.  \citet[p. 377-8]{SteinGrunbaum}, meanwhile, emphasizes the history of the development of geometrical ideas in physics, noting that the idea that spacetime \emph{has} a geometry in the first place, in the modern sense, arises from studying the dynamics of matter understood to ``probe'' that geometry.}

To be sure, even if one takes general relativity to make certain claims regarding ``spacetime geometry'' in this stronger sense, so that it implies that matter dynamics is adapted to that geometry, one still has to, at some stage or other, address the issue of what dynamics are ``allowed'', in the sense of adverting to just the relevant geometry.  But now one understands those requirements to be a clarification of what one means by asserting that spacetime has some geometry in the first place.  The mere fact that someone has written down a field with certain mathematical properties should not be taken to have any consequences for the dynamics of matter or ``chronogeometry'', as the dynamicist sometimes claims the geometricist believes.  What matters is that it is the intended physical significance of that field that it does have those consequences. Thus, on this view of spacetime geometry, matter that does not have the right sort of dynamics is simply incompatible with general relativity.\footnote{An anonymous referee raises the following worry: is it not the case that physicists often impose ``extra'' constraints on matter fields, such as energy conditions, which might be understood as limiting attention to just some of the many matter fields that are prima facie compatible with relativity theory (but which we reject on other grounds)?  The answer is complicated by the different energy conditions in the literature and the various roles they play.  But I would argue (a) at least in some cases, some energy conditions (such as the dominant energy condition---see note \ref{DEC}) serve as a guide to whether a proposed matter dynamics is suitably compatible with general relativity (e.g., Maxwell's equations are compatible, in part because solutions always satisfy the dominant energy condition relative to whatever metric appears in the equations and stress-energy tensor; but arguably tachyon fields are not, because this condition fails), so that they are not ``imposed'' so much as ``checked''; (b) in other cases, such as in proving singularity theorems, one uses energy conditions as a stand-in assumption, capturing the idea that one is considering only ``reasonable'' matter (which may be glossed as: matter compatible with the theory), so that although one has to impose a condition, doing so reflects precisely the relationship between matter and geometry described here; and (c) in yet other cases, energy conditions are introduced for technical convenience but have obscure physical motivation.  This list may not be exhaustive, but I believe it covers the main cases.}

All that said, at this point the debate begins to seem terminological.  Once we agree that (d) is necessary, and the issue becomes only whether (g) should be understood in such a way as to subsume (d) or whether it is better to state (d) separately (just to be safe), it is hard to see much of substance left to the disagreement.\footnote{This is not to say that there have not been other disagreements of substance adjacent to the one discussed here.  For instance, there was a real disagreement, now resolved, regarding whether in general relativity, the geodesic principle (discussed in what follows) is a ``consequence of Einstein's equation''; for more on this, see \citet{Brown}, \citet{MalamentGP}, and \citet{WeatherallGP}.  There is also an unresolved (I believe) disagreement about whether there is some salient difference between general relativity and other theories, such as Newtonian gravitation, regarding the status of inertial motion \citep{WeatherallGP, Sus, WeatherallPuzzleball}.  Finally, there have been disputes about whether spacetime geometry can provide ``constructive explanations'' \citep{Brown+Pooley1,Brown+Pooley2,Janssen,Dorato,Frisch,Acuna,ReadConstructive}; and about whether one can even state matter dynamics without specifying some geometrical background \citep{Norton,WallaceWACS}.}

This conclusion substantially recapitulates \citet{ReadIreny}.  Still, he puts the point somewhat differently.  As noted above, he suggests that there are two versions of the geometric view.  One of these, which he calls the ``unqualified geometrical approach'' (UGA), maintains that (d) follows from (g), but apparently interprets (g) in a weak way, such that it is hard to see how the defender of this view could possibly get from (g) to (d)---and \citet{ReadIreny}, following \citet{Brown} and \citet{Miracles}, offer numerous examples of theories that seem to show you cannot.  The other geometrical view, which he calls the ``qualified geometrical approach'' (QGA), meanwhile, supplements (g), or understands it in a stronger way, such that one assumes (for instance) that the metric and its associated Levi-Civita derivative enter into the matter dynamics in certain ways, so that matter dynamics are assumed, or required, to be adapted to the spacetime geometry. (It is the QGA that I have called the geometric view above.)  Here, Read argues, (d) does follow from (g) (or, trivially, from (g) and (d) together); and in some cases one might even concede that (g) \emph{explains} (d).  It is the QGA that Read suggests is indistinguishable from the dynamical view as he understands it.

\section{A Little Too Irenic}

So much for (I)---and for agreement.  I will now turn to (II).  Let us stipulate that we require both (g) and (d).  Even so, it would be desirable to have some precision and clarity on what (d) means.  In particular, what sorts of matter dynamics are compatible with (d)?

There are really two separate questions that one might like to answer here.  First, one would like to know if we can identify:
\begin{enumerate}[start=1,label={(\Alph*)}]
\item Features that ``adapted'' dynamics will (necessarily?) have.
\end{enumerate}
Some generally accepted candidates for (A)---the hallmarks of adapted dynamics---are often thought of as ``principles'' of general relativity: the dynamics should imply, or at least be consistent with, the idea that small massive bodies follow timelike geodesics (the ``geodesic principle''); that certain kinds of matter, in an ``optical'' or ``ray'' limit, follow null geodesics; that clocks record arclength (or ``proper time'') along their trajectories (the ``clock hypothesis''); that various energy conditions are satisfied; and so on.  But it is often difficult to even state these conditions clearly and precisely, at least in a way that makes any direct connection to the sorts of systems of differential equations (e.g., Maxwell's equations) that express matter dynamics in general relativity.  Thus it is not entirely clear how (A) is supposed to be related to (d).  To address this gap, one might also wish to identify:
\begin{enumerate}[start=2,label={(\Alph*)}]
\item Necessary and/or sufficient conditions for a system of partial differential equations to exhibit the features of ``adapted dynamics''.
\end{enumerate}
In other words, with (B) we are looking for conditions on matter dynamics such that the features identified in (A) obtain.

To this end, there are various heuristics available for constructing matter theories, in the sense of systems of partial differential equations, that one expects to be adapted to a given spacetime geometry.  These heuristics go by names such as ``minimal coupling'', ``the comma to semicolon rule'', etc.  But these heuristics are sometimes ambiguous, and in any case, it is not clear that they work, in the sense that it is not clear that these heuristics necessarily generate theories that exhibit the features (A) that we require for dynamics to be ``adapted to the geometry''.  What one would really like, in my view, would be theorems of the following form
\begin{quote} Let $(M,g_{ab})$ be a relativistic spacetime. Any system of partial differential equations on $M$, with such and such further properties [e.g., identified via one of the heuristics just notes] will be adapted to the geometry of $(M,g_{ab})$''.
\end{quote}
Of course, this is not a precise statement, both because the antecedent has not been specified and because the consequent is not precise.  It is also not clear that one wants a single theorem, here; more likely is that there will be a network of results, demonstrating the ways in which different senses of ``being adapted to the geometry'' imply and constrain one another.\footnote{I explore this idea, that the principles of a theory should be understood to form a network of logical interdependencies, in \citep{WeatherallPuzzleball}, where I call it ``the puzzleball conjecture''.}

One route to making this sort of schematic statement more precise is to use small body motion as a candidate for (A).  The idea, here, is that the geodesic principle, which states that in the absence of external forces, free massive test bodies traverse timelike geodesics of the spacetime metric, provides a link between ``geometrically privileged'' curves---that is, the geodesics, a class of curves picked out by the metric---and ``physically privileged'' curves---namely, the trajectories of small, force-free bodies.  A similar link is provided by the ``light ray principle'', that light rays follow null geodesics.  These links suggest a candidate interpretation for at least part of what it means for matter dynamics to be ``adapted to the geometry'': namely, differential equations are ``adapted to the geometry'' (only) if the solutions of those equations, in some appropriate small-body limit, (only) follow (timelike or null) geodesics.  One might then try to identify generic conditions on matter equations (i.e., candidates for (B)) that allow one to conclude that matter \emph{is} adapted to the geometry in this way.

And of course, as has been widely discussed in the recent philosophical literature, such theorems exist.  For instance, the main result of \citep{Geroch+Weatherall} may be put as follows.\footnote{See \citep{WeatherallGM} for a discussion of these results aimed at philosophers.}   Fix a spacetime $(M,g_{ab})$, and let $\nabla$ be the Levi-Civita derivative operator associated with $g_{ab}$.  Suppose that $\mathcal{C}$ is a collection of smooth, symmetric rank two tensor fields---say, the stress-energy tensors associated with the solutions to some matter field equations---and suppose that those fields are all divergence-free with respect to $\nabla$ and that they all satisfy a certain energy condition with respect to $g_{ab}$.\footnote{\label{DEC} Specifically, Geroch and Weatherall require the \emph{dominant energy condition}, which holds of a tensor $T^{ab}$ at a point if, for any pair $\xi^a,\eta^a$ of timelike vectors there, $T^{ab}\xi_a\eta_b\geq 0$.}   Then the collection $\mathcal{C}$ \emph{tracks} (only) timelike or null geodesics.  This, in turn, implies that these are the only curves around which ``small-body-like'' solutions can propagate.\footnote{Note that although tracking has this consequence, one might also take tracking---for the details of which I refer the reader to \citep{Geroch+Weatherall}---to provide a direct link between the solutions of a system of partial differential equations and curves that does not need to go through small bodies.  So tracking could itself count as a partial explication of being ``adapted to the geometry''.  See \citep{WeatherallGM} for more on this idea.}

Now, as I emphasize in \citep{WeatherallGM}, the stress-energy tensors associated with solutions to standard ``relativistic'' equations, such as the source-free Maxwell's equations and the Klein-Gordon equation (but not the Dirac equation), necessarily satisfy the two conditions stated.  Moreover, as I note in \citep{WeatherallConservation}, drawing on well-known arguments \citep[e.g. in][appendix E]{Wald}, the stress-energy tensor associated with any matter field will be divergence-free, thereby satisfying one of the two conditions, as long as its dynamics are determined by a Lagrangian with certain properties---basically, one in which the spacetime metric and associated derivative operator play certain roles, in a sense I make precise there.

One might ask for still more.  For instance, it would be very nice to have a general argument capturing necessary conditions for the dominant energy condition, which is needed for the Geroch-Weatherall theorem, to be satisfied by the stress-energy tensors associated with arbitrary solutions to a differential equation, analogous to the conditions under which that tensor is guaranteed to be divergence-free.  One imagines such results are possible, though the details are not perfectly clear, in part because it is known that solutions to some ``relativistic'' equations, such as the Dirac equation and some scalar field equations, do not necessarily have this property.\footnote{\citet{CurielEC} provides a general discussion of cases where various energy conditions are known to fail.}  Does this mean that they are not ``adapted to the geometry''?  Or does it mean that we need a broader understanding of that expression?

One might also ask whether there are other results in this neighborhood that would capture a more expansive notion of what it means for matter to be adapted to spacetime geometry.  For instance, could one capture something like the ``clock hypothesis'', which states that suitably ideal clocks measure arclength along their worldlines?  \citet{FletcherCH} makes some progress along these lines, but it relies on a particular construction (that of the ``light clock'');\footnote{To be clear, Fletcher is not committed to using his result for the purposes I sketch here, and so what I write in the main text should not be taken as a criticism of his result.} one might ask for something more general, such as a result to the effect that any ``periodic'' behavior, suitably defined, associated with solutions to matter field equations satisfying certain conditions (candidates for (B)), measures arclength in a suitable small body limit.  I do not know if such results can be achieved, but they would be a further elaboration of this research project.  One could also try to show that such matter does not propagate superluminally, for instance in the sense of ``signal velocity'' defined by \citet{GerochFTL}.\footnote{\citet{EarmanSL} and \citet{WeatherallSL} also discuss this definition, including ways in which it does and does not align with, for instance, the dominant energy condition and other senses in which matter might be said to be ``relativistic''.}

The line of research I have just described is one, still incomplete, approach to answering (II).  But it is apparently not the one favored by the Brown and collaborators.  And here, I think, is a point where there remains potentially significant disagreement between the ``dynamical school''---that is, the avowed defenders of the dynamical view---and others.\footnote{I say ``dynamical school'' because I do not believe that there is an essential link between the program I will presently describe and the dynamical view as I have presented it, viz., the idea that (d) is independent of (g) on a weak reading of (g).  And yet defenses of the dynamical view seem invariably to invoke the ``strong equivalence principle'' or similar ideas in what I take to be a problematic way.}  In particular, \citet[pp. 15-6]{Miracles}, following \citet[pp. 169-171; 176]{Brown}, offer a different analysis of (d).  Those authors apparently assert that the so-called ``strong equivalence principle'' (SEP) is necessary (and also sufficient?) for (d) to hold.\footnote{\citet[p. 10-11]{ReadIreny} writes that the SEP is an ``important condition for the chronogeometricity of the metric field'' (p. 10), but hedges on whether it should be viewed as strictly necessary, sufficient, or jointly sufficient along with some other assumptions.  In more recent correspondence, Read has suggested his views have shifted on this issue, and that he now sees the SEP primarily as establishing a link between general relativity and special relativity.}

Here the strong equivalence principle is understood as the following statement:\footnote{The status of the SEP is discussed in some detail in a very nice paper by \citet{Misconceptions}; see also \citet{Miracles} and \citet{Fletcher}.}
\begin{quote}\singlespacing
There exists in a neighborhood of each event preferred coordinates, called \emph{locally inertial} at that event.  For each fundamental non-gravitational interaction, to the extent that tidal gravitational forces can be ignored, the laws governing the interaction find their simplest form in these coordinates.  This is their \emph{special relativistic form}, independent of spacetime location. \citep[p. 169]{Brown}
\end{quote}
On Read's presentation, at least, the idea is that we should understand general relativity to be a theory (and here I am switching to Read's own way of presenting things) whose ``dynamically possible'' models consist of a Lorentzian manifold $(M,g_{ab})$, along with other fields representing matter, such that (i) the matter fields (and their interactions) are associated with stress-energy tensors, and the metric $g_{ab}$ satisfies Einstein's equation with the sum of these stress-energy tensors as its source; and (ii) the matter fields satisfy certain differential equations that are compatible with the SEP.  In other words, the SEP is a constraint on matter fields; and if it is not satisfied, then we do not have a truly ``generally relativistic'' theory.  It is for this reason that I take satisfaction of the SEP to be a proposal concerning precisely what is meant by (d).

I would like to raise two concerns about the proposal that the SEP, as just stated, is a satisfactory way of making precise the idea that ``matter dynamics are adapted to spacetime geometry''.\footnote{There is a long history of controversy over the equivalence principle in its various guises, where semi-precise statements are offered and then shown to be inadequate for one reason or another.  I do not mean to engage in this debate here: my goal is not to argue that Brown or Read has failed to capture what Einstein had in mind when he formulated his principle, nor do I wish to claim that this or any other version of the equivalence principle is trivial or false.  I am specifically interested in the question of whether the assertion that I have called the strong equivalence principle, as elaborated and elucidated by Read, adequately captures the idea that matter dynamics are ``adapted to spacetime geometry''.  Moreover, although I do mean to criticize one aspect of how Brown and Read have used the SEP, it is hardly as if Brown, much less Read, has invented this principle: it is common fare in textbooks on general relativity \citep[e.g.][p. 386]{MTW}. The concerns I am raising are as much directed at these classic textbook treatments as at the contemporary philosophers who have invoked them.}  The first concern is that it is not clear what the strong equivalence principle says.\footnote{The first concern is similar in several important respects to arguments made by \citet{Fletcher}; I did not see that manuscript until this one was drafted, but my perspective was certainly shaped by conversations with him.}   And the second concern is that it is not clear that the SEP is sufficient.

\subsection{The SEP is Not Clear}

What is unclear about the SEP?  There are several issues.  One, which Brown, Read, and others acknowledge, is that it is not clear what it means to say ``to the extent that tidal gravitational forces can be ignored''.  Here is just one reason one might be worried about this issue. The SEP, as a foundational principle, is supposed to hold in all regimes. But then how are we to understand ``strong field'' regimes, where curvature, by various natural measures, is large, or unbounded along a curve?  To what extent can tidal forces, or curvature more generally, be ignored in these regimes---such as when the components of the curvature tensor, expressed in ``locally inertial'' coordinates, are larger than any other quantity under consideration?  One presumably would like to say that the SEP holds \emph{approximately} in all cases, but this notion of approximation is subtle, and it has not been clearly laid out. As I say, however, this concern has been raised previously, so I will not focus on it here; just suffice it to say that until one is clear about what this proviso means, it is very difficult to establish what equations do in fact satisfy the SEP, or what the consequences of the SEP are.

There is another issue, however, which I do not think has received adequate discussion in the literature.\footnote{My framing of this issue is particularly indebted to conversations with Sam Fletcher; a number of closely related issues are addressed in \citet{Fletcher} and \citet{Fletcher+Weatherall}.}  It concerns the role that the expression ``simplest form'' plays in Brown's version of the SEP.\footnote{There is a third issue here that I do not emphasize, which is that it is not clear why special relativity is given primacy of place a century after general relativity superseded it.  In particular, when one introduces a new matter theory---say, a theory of an inflaton field in the early universe---why should one be obligated to first identify a ``special relativistic'' form for the dynamical equations of that theory?  One possible answer---and I am grateful to an anonymous referee for raising this---is that the Brown-Pooley constructive/dynamical relativity program was originally developed in the context of Minkowski spacetime, and that Brown then aims to extend it to general relativity. (This observation is perhaps connected to Read's position that the dynamical and geometric approaches \emph{are} distinct in the context of special relativity---but collapse once one moves to general relativity.)  From this perspective, it arguably does make sense to interpret general relativity through the lens of an already-worked-out position on special relativity.  But even so, I think it would be preferable if one could do the work the SEP is meant to do, at least for establishing (d) in general relativity, without using a different theory as a crutch.  Moreover, the approach to (d) that I defend above, involving results such as those of \citep{Geroch+Weatherall}, works just as well for special relativity as for general relativity, since any general result will hold in flat spacetime as a special case.}  As a first pass at the worry, I claim that it is not clear how to evaluate whether an equation is in its ``simplest'' form: it seems background assumptions about simplicity and conventions about how to express equations must go into such judgments.\footnote{\citet{Miracles} suggest that the simplest form of an equation is one in which no terms vanish by coordinate transformation, but I find this account inadequate for reasons that I hope will be clear presently.  See note \ref{miraclesRV}.}  For instance, it is not clear why one should expect the simplest form of any equation relating structures that can be characterized in a coordinate \emph{independent} manner to be a representation in certain coordinates; one might even think that insofar as any coordinate representation of an equation invokes some further structure---namely, a choice of coordinates---it is \emph{ipso facto} less simple than a coordinate independent expression of the same equation.  (Of course, this is not what Brown has in mind; he means something like the ``simplest form among all coordinate expressions''.)  To be sure, the issue is not that Brown requires that the form be \emph{simple}, per se, or that he implicitly adopts a particular account of simplicity that I reject.  Rather, the issue is that Brown wishes to use something about the form of the equations---viz. the simplicity of that form---as a signal that those equations are appropriately adapted to certain coordinate systems or other structure.  The trouble is that the ``simplicity'' of the form of an equation is not a sufficiently clear or unambiguous notion to play this role, at least without further work.

But there is a deeper worry lurking here. The thing whose simplicity is being evaluated, on this approach, is the ``form'' of the equation, i.e., its syntactic expression, relative to certain (unspecified) conventions concerning how to ``express'' an equation in certain coordinates.  But one might have expected, from the very fact that equations \emph{can} take more or less simple forms, that ``equations'', properly construed as mathematical objects, do not stand in one-to-one correspondence with expressions or ``forms'' for those equations.\footnote{This is just a reflection of the well-known fact that propositions can be given many different provably equivalent syntactic forms, e.g., $p$, $p\wedge \top$, $p\wedge \bot$, etc., some of which are simpler than others (irrespective of any choice of coordinates).  \citet{Fletcher} raises a similar concern with minimal coupling, arguing that it is ``hyperintensional'' in the sense that it distinguishes between cases that ought to be seen as equivalent.}  And indeed, this is true.  Even within a single coordinate system, there may be multiple different ways of expressing the same ``equation'', in the sense of expressing a relationship between variables that is satisfied by the same valuations of those variables, some of which are simpler than others even within a coordinate system; and some of these ways of expressing an equation may take apparently \emph{simpler} forms in \emph{different} coordinates.

These are abstract claims, so let me illustrate with an example.  Fix a spacetime $(M,g_{ab})$.  We will say (following Brown) that coordinates in a neighborhood of a point $p\in M$ are ``locally inertial at $p$ relative to $g_{ab}$'', or just ``adapted to $g_{ab}$'', if $g_{ab}$, expressed in those coordinates, is diagonal and normalized at $p$; and the Christoffel symbols for the Levi-Civita derivative operator associated with $g_{ab}$ vanish at $p$.  Such coordinates can always be found at any point, for any Lorentzian metric.  Now consider, for instance, the following system of equations in Minkowski spacetime, $(M,\eta_{ab})$:
\begin{subequations}\label{EqMaxNew}
\begin{align}
\tilde{\nabla}_aF^{ab} &= \mathbf{0}\\
\tilde{\nabla}_{[a} F_{bc]}=\mathbf{0},
\end{align}
\end{subequations}
where $\tilde{\nabla}$ is the unique derivative operator such that $\tilde{\nabla}_a(\Omega^2 \eta_{bc})=\mathbf{0}$ for some (fixed) non-vanishing, non-constant smooth scalar field $\Omega$.\footnote{Here $[\cdot]$ indicates antisymmetrization on indices; and I raise and lower indices using $\eta_{ab}$, the Minkowski metric.}

Do these equations satisfy the SEP?  To answer this, we need to determine whether these equations take their ``simplest form'' in locally inertial coordinate systems relative to $\eta_{ab}$.  I believe, based on how Brown approaches similar examples, that the answer should be ``no''.  These equations take their simplest form in coordinate systems that are locally inertial relative to the metric $\Omega^2\eta_{ab}$, not ones that are locally inertial relative to $\eta_{ab}$.  As evidence for this, observe that, expressed in coordinates adapted to $\eta_{ab}$, the Christoffel symbols that appear in the first of these equations will not vanish, while they will vanish in coordinates locally inertial relative to $\Omega^2\eta_{ab}$.  It appears to follow that these equations do not satisfy the SEP.  But this would be a surprising conclusion, since after all these are simply Maxwell's equations in Minkowski spacetime---that is, they determine the same solutions---where I have written them in a form that exploits the well-known conformal invariance of those equations.\footnote{One might worry that Eqs. \ref{EqMaxNew} are not, after all, equations ``in'' Minkowski spacetime---say, because the derivative operator and metric appearing in the equations are not the Minkowski metric and derivative operator, or because they do not take their ``simplest form'' in coordinates adapted to $\eta_{ab}$---even though they have precisely the same solutions.  For my own part, I do not think this is a compelling position, because I think the most natural sense, mathematically, of ``same (differential) equation'' is ``has the same solutions''.  But if one wishes to argue that the cogency of the SEP depends on adopting a \emph{different} standard of equivalence for equations, that route is certainly available---but the very fact that there could be disagreements on this point shows that the SEP is unclear in just the way I am claiming, because the class of manipulations that preserve an equation have not been specified.}

One might respond that Maxwell's equations take ``equally simple'' forms not only in the coordinates adapted to $\eta_{ab}$, but in others as well; this is compatible with the SEP, since after all, the SEP does not say that these equations take their simplest forms \emph{only} in locally inertial frames.  This is a fair observation, but it is not to the point, which is that it is not clear how to tell which form is to count as the simplest, or what operations are allowed to simplify them.\footnote{\label{miraclesRV} This argument is why I do not take the claim that equations are in their simplest form if terms do not drop out by changing coordinates to be satisfactory, because there may be other transformations available that further simplify equations.}  Put another way: defenders of the SEP have not provided a clear criterion of individuation for equations, such that one can say when two different strings correspond to a single equation in different forms, as opposed to different equations. In the present case, it turned out that there were additional operations, beyond simply expressing the equations in a particular coordinate system, that could be performed on the equations while preserving a crucial feature of those equations---that is, their class of solutions---to simplify them further within some coordinates.  How were we to know these were available (especially since they are not available for all equations, such as the Klein-Gordon equation)?  The answer is by studying the transformations that preserve something other than the form of the equation---specifically, its solutions.  I take this to mean that what is---or should be---at issue in the SEP is not really the ``form'' of an equation at all, but rather properties of the equation that are intrinsic to the equation, i.e., that are invariant under different forms the equation might take.  That the equation takes a certain form, with certain conventions, using certain background knowledge about the equation, is at best an indirect guide to those properties and at worst too ambiguous to be useable.

How might one get at these properties of equations?  One strategy would be to look to symmetry properties of the equations as a way of identifying the relevant ``intrinsic'' structure of those equations, independent of particular representational choices.  Read, following \citep{Miracles}, proposes a further clarification of the SEP that appears to adopt this strategy.\footnote{It is not clear that Read's motivation for using a symmetry condition is to avoid possible ambiguities arising from the notion of the ``form'' of an equation; the motivation offered in \citep{Miracles}, for instance, is the concern that curvature terms may appear in some equations that ``should'' satisfy the SEP.}  He writes (after setting aside issues about ``neglecting gravitational forces''):
\begin{quote}\singlespacing
For today, the essential aspect of the SEP is the imposition that, in the neighborhood of any $p\in M$ in GR, laws of physics recover their `special relativistic form'---where I shall understand this to mean: a \emph{Poincar\'e invariant form}.  Clearly, this is a particular restriction on the matter sector in the theory. \citep[p. 9]{ReadIreny}
\end{quote}
Here having a ``Poincar\'e invariant form'' is replacing the idea of having a ``special relativistic form'' (or a ``simplest form'').  The same concepts appear in Read and collaborators' discussions of the ``miracles'' of relativity, which may similarly be read as an elaboration of the SEP.
\begin{quote}\singlespacing \textbf{MR1}: All non-gravitational interactions are locally governed by Poincar\'e invariant dynamical laws. \citep[p. 14]{ReadIreny}\end{quote}
\begin{quote}\singlespacing \textbf{MR2}: The Poincar\'e symmetries of the dynamical laws governing non-gravitational fields in the neighbourhood of any point $p\in M$ coincide (in the regime in which `gravitational forces' can be ignored) with the symmetries of the metric field in that neighborhood. \citep[pp. 14-5]{ReadIreny}\end{quote}
In other words, it is not merely that all of the laws are required to have a certain symmetry property; it is that those symmetry properties must be appropriately coordinated, both across different matter fields and with the spacetime metric.

The invocation of the ``symmetries of the \emph{metric} in [a] neighborhood'' (emphasis added) in the statement of the second miracle suggests that Read might have in mind a standard definition of the ``local symmetries'' of a metric, which associates such symmetries with the existence of Killing fields in a neighborhood of a point; this notion of local symmetry might then be extended to capture the idea of a ``local symmetry'' of an equation.\footnote{Recall that a Killing field of a metric $g_{ab}$ is a vector field $\xi^a$ satisfying Killing's equation: $\nabla_{(a}\xi_{b)}=\mathbf{0}$, where $\nabla$ is the Levi-Civita derivative operator associated with $g_{ab}$.  Killing's equation implies that the Lie derivative of $g_{ab}$ along $\xi^a$ vanishes, which in turn means that there exists locally a one parameter family of isometries generated by $\xi^a$.  These isometries are ``local symmetries'' in the straightforward sense that they are local maps that preserve the metric.  See \citet[\S 1.9]{MalamentGR} for more details on Killing fields.  }  But this notion of ``local symmetry'' cannot be quite what Read is looking for.\footnote{\citet[\S 2.2]{ReadIreny} discusses the notion of symmetry that he does have in mind, and appears to identify it with ``isometries'', which would suggest that Killing fields \emph{are} the generators of local symmetries in his sense; he goes on to acknowledge that the ``the metric field $g_{ab}$ in GR need not in general gave any non-trivial symmetries'' (p. 6).  This shows, I think, that there are internal tensions in his discussion of the miracles that bear on the SEP, and which are worth discussing.}    The difficulty concerns the role that the proviso ``in the regime in which `gravitational forces' can be ignored'' plays in the second miracle, where these local symmetries are invoked. In a general spacetime $(M,g_{ab})$, if the curvature tensor is non-vanishing at a point, or even if it vanishes at that point but is non-vanishing in every neighborhood of that point, the metric will in general have \emph{no} ``local symmetries'', in the sense of having no Killing fields.  (Conversely, if the curvature tensor \emph{does} vanish in a neighborhood of a point, the metric \emph{does} have Killing fields in that neighborhood.)  It is not clear how assuming that curvature is ``sufficiently small'' or ``locally negligible'' avoids this issue; and it means that the second miracle becomes trivial in the context of curved spacetimes.\footnote{There is another problem with this condition, which is that even if cases, with curvature, where one \emph{does} have a Killing field, in general those fields will not correspond to ``Poincar\'e symmetries''.}

One might reply by saying ``the second miracle has bite precisely in those cases where the metric \emph{does} have local symmetries; and the SEP rules out equations that, in highly symmetric cases (say, in flat spacetime), have symmetries that do not coincide with the metric symmetries''.  But this does not seem like an option that is open to Read, because the first miracle explicitly states that the ``local symmetries'' of the dynamical laws are supposed to (always) include the Poincar\'e group, while the second miracle states that these are supposed to coincide with the local metric symmetries, and in general the Killing fields of a metric at a point do \emph{not} include the (generators of the) Poincar\'e group.

Besides, even if one were to weaken the statement of the first miracle, focusing on cases in which the local symmetries of the metric \emph{are} the (generators of) the Poincar\'e group (i.e., focus on flat spacetime), it would not help.  Consider the equation $\nabla_a\nabla^a\varphi + \xi R\varphi=0$ on a spacetime $(M,g_{ab})$, where $R$ is the curvature scalar and $\xi\neq 0$ is some number.  In any neighborhood where curvature vanishes identically, this equation will reduce to the wave equation, and presumably will satisfy the SEP (if we use the miracles as a guide).\footnote{Likewise, if one has Killing fields in a neighborhood of a point, then the one parameter families of isometries generated by those Killing fields will be ``symmetries'' of this equation, in the natural sense that for any solution $\varphi$ in that neighborhood, the pullback of $\varphi$ along each isometry in the family will also solve the equation in an appropriate neighborhood.}  But this is presumably the wrong verdict, since if one were to express this equation in coordinates in which the metric is diagonal at a point and all Christoffel symbols vanish at that point, a curvature term would still appear, which I would think means that the equation is not in a ``special relativistic form''.  (On the other hand, perhaps we are supposed to ignore this curvature term, since we are assumed to be operating in ``in the regime in which `gravitational forces' can be ignored''.)\footnote{Given the analyses in \citet{Miracles} and \citet{Misconceptions}, one might think that curvature terms are compatible with the SEP, properly construed, as long as they appear only in second order equations.  One might take this to mean that various ``non-minimally coupled'' scalar fields should count as satisfying the SEP.  But then consider the (first-order) equation $\nabla_a\xi^a + R\xi^a\xi_a = 0$.  The same argument applies.}

One might also reply by saying that I am placing too much emphasis on \emph{exact} symmetries; really, what Read et al. need is a notion of ``approximate'' Poincar\'e symmetry, where the degree of approximation improves as curvature approaches zero.  The idea of an ``approximate symmetry'' is notoriously difficult to make precise, however, and seems to require reference to some background structure.  For instance, \citet{Fletcher} proposes a natural, but frame-dependent, definition of ``approximate (Poincar\'e) symmetries'', which may be extended to a notion of ``approximate Killing fields''.  Here, the notion of ``approximate symmetry'' at issue is one in which the change in the spacetime metric at a point as one flows along a vector field, as determined by a positive definite metric (given by a choice of frame field) is less than some choice of $\epsilon$, for sufficiently small values of the flow parameter.

Could Fletcher's proposal help Brown and Read?  At first glance the answer may seem to be ``yes''.  In particular, Fletcher shows that every Lorentzian manifold has approximate Poincar\'e symmetries, to any degree of approximation, in sufficiently small neighborhoods of every point.  He also shows that the approximate local Poincar\'e symmetries of a metric can come apart from the Poincar\'e symmetries of a system of equations, in the sense that there exist examples of equations in a relativistic spacetime that do not exhibit local Poincar\'e invariance (at least on one construal of what it means for equations to have such a property)---even though all spacetimes have local approximate Poincar\'e symmetry.\footnote{On the other hand, Fletcher's example is that of a dust field in Minkowski spacetime, which is not ``locally Poincar\'e invariant'' because its equations take an especially ``simple form'' in comoving coordinates.  That this example might fail to satisfy the SEP suggests that something has gone badly wrong in the whole program.  But I set this issue aside, because I think it is ultimately a red herring for present purposes.}  This claim seems compatible with, and even supportive of, the idea that it is ``miraculous'' that these symmetries might be said to coincide for the equations governing realistic matter.  But on further reflection, I do not think the approximate Poincar\'e symmetries Fletcher considers could possibly play the requisite role in the SEP.  The reason is that the notion of ``approximate local symmetry'' that Fletcher defines is far too permissive.  In particular, \emph{every} vector field on a manifold $M$ is an approximate local symmetry, in sufficiently small neighborhoods of any point $p\in M$, for any smooth metric on $M$ at all---purely by continuity considerations.  Hence, although there is always some representation of the Poincar\'e algebra (i.e., the generators of the Poincar\'e group) in the vector fields in a neighborhood of any point such that it is true that any Lorentzian metric has ``local approximate Poincar\'e invariance'', that representation is not unique, and moreover, the same would hold for many other Lie algebras as well.  Thus, on this understanding of ``local approximate symmetry'', there is a strong sense in which the second miracle would trivialize.

Having now considered what Read (and Brown) apparently do not mean by ``local (Poincar\'e) symmetry'' (but which one might have thought they would mean), I will consider what it seems they do have in mind. Here Appendix A to \citep{Miracles} is especially helpful.\footnote{I take the manipulations there to be consistent with the definitions given in \citep[\S 2.2]{ReadIreny}: ``A coordinate transformation is a \emph{dynamical symmetry} just in case the dynamical equations governing non-gravitational fields take the same form in coordinate systems related by that transformation'' (pp.6-7).  \citet{Dewar} also engages in detail with this appendix.} There, the authors of that paper present an argument that ``minimally coupled dynamical equations in GR manifest local Poincar\'e invariance''.  They proceed by fixing coordinates in a neighborhood of a point $p$; expressing a given equation in that neighborhood in those coordinates (including expressing derivatives and curvature tensors using partial derivatives and Christoffel symbols); and then showing that the syntactic expression of each term in this equation does not change under certain coordinate transformations---namely, those that are ``relatively constant'' (i.e., have the same domain and give rise to the same partial derivative operator) and which do not change the expression of the metric at $p$.   Terms (i.e., summands) in an equation whose expression is unchanged by this procedure are said to be ``locally Poincar\'e invariant''; an equation is ``locally Poincar\'e invariant'' if each of its terms is.\footnote{\label{construal} Here is a stab at a more precise, less ``syntactic'' statement. One can always find, in a sufficiently small neighborhood of $p$, a flat metric $\eta_{ab}$ that agrees with $g_{ab}$ at $p$, and whose Levi-Civita derivative operator $\bar{\nabla}$ agrees with that of $g_{ab}$ at $p$.  This metric is essentially unique, in the sense that any two such metrics are related, within a sufficiently small neighborhood of $p$, by an isometry that leaves $p$ fixed.  Rewrite derivatives in your equation in terms of $\bar{\nabla}$ and a tensor field $C^a{}_{bc}$. Say a geometrical object $X$ (i.e., one for which a pushforward is defined) is ``locally Poincar\'e invariant'' (relative to $\eta_{ab}$) if for any smooth, smoothly invertible map $\varphi$ from a sufficiently small neighborhood of $p$ to itself, which acts as an isometry for $\eta_{ab}$ and leaves $p$ fixed, is such that $\varphi_*(X)_{|p}=X_{|p}$.  Note that on this recovery, ``Poincar\'e'' seems like a misnomer, since we do not consider translations.}

It is clear from this that, by construction, any Lorentzian metric is ``locally Poincar\'e invariant''; but it is equally clear, as has already been suggested, that this notion of ``local symmetry'' has nothing to do with Killing fields or local isometries, nor is it really a notion of symmetry ``in a neighborhood'', so much as ``at a point''.  Note, too, that this notion of symmetry is also not a candidate notion for ``dynamical symmetry'' in anything like the standard sense of that term, because it does not relate solutions of a differential equation to other solutions of the equation; it just relates ``expressions'' to ``expressions'', and does so only at a point.\footnote{Moreover, if one recovers the condition as in the previous footnote, in general ``local Poincar\'e transformations'' will not take solutions to solutions for standard equations, including, e.g., Maxwell's equations in curved spacetime.}  Finally, on this interpretation of what Brown and Read have in mind, it would appear that equations, including equations that are first order in a matter field, that include curvature terms, may satisfy the SEP.  Indeed, the construction is such that \emph{any} equation whose terms are (non-zero) vectors will fail to be Poincar\'e invariant; whereas any scalar term will automatically count as Poincar\'e invariant, even if that scalar results as the inner product (say) of vectors or tensors.\footnote{Compare the claim here that any equation whose summands are vectors cannot be Poincar\'e invariant, which holds simply because the components of any non-zero vector will always change when one applies a Lorentz boost or rotation, to the discussion surrounding \citep[Eq. A.12]{Miracles}, where the authors apparently claim that an equation with terms of just this character \emph{is} Poincar\'e invariant.  I believe that their claim is a slip, and that their argument turns on an ambiguity about whether the sense of ``same form in a coordinate system'' is supposed to mean ``same components'' (which is how I have interpreted them here) or ``same symbolic representation'', which I fear is trivial, since one is free to use whatever symbols one likes.}  But perhaps the most important point is just this: although in moving from ``simplest form'' to ``Poincar\'e invariant form'', Read and collaborators invoke a symmetry principle in their new definitions, it still concerns purely syntactic information, namely, how an expression looks when written according to a particular convention.  And so the worries expressed above continue to hold.

I have presented a number of arguments in this subsection, all to the effect that the SEP is not sufficiently clear, so that one cannot unambiguously determine whether it holds of a given system of equations.  In the next subsection, I will present another argument against the SEP, as formulated above, which I take to be ultimately more important.  But before proceeding, it may be valuable to briefly summarize what has gone so far.  The concerns I have raised may be condensed into four main points: First, as has been noted by others, it is not clear what it means to say ``to the extent that tidal forces can be ignored''; second, it is not clear how to make precise the notion of ``simplest form'' of a system of equations, and different informal understandings of this notion seem to yield different conclusions; third clear identity criteria for equations have not been given, so that one cannot determine when one has a single system of equations expressed in multiple forms (some of which may be simpler than others) as opposed to distinct equations, and thus it is not clear what manipulations of equations are permitted when trying to determine their simplest form; and finally, fourth, a precise notion of ``(approximate) local symmetry'' that captures a non-trivial relationship between the symmetries of an equation and the symmetries of a metric has not been given, at least in curved spacetime.

\subsection{The SEP is not Sufficient}

I will now turn to the second concern.  Suppose the difficulties just noted can be overcome and we have a clear and precise formulation of the SEP, such that one can, without ambiguity, determine whether a given system of differential equations satisfies it; and it expresses something about the ``instrinsic'' character of a system of equations, and not their expression.  (I think this is very likely possible, though it would involve introducing different methods from those usually employed in this literature.)  Then as in the arguments offered above, one would presumably expect it to follow that such a system is, indeed, ``adapted to the spacetime geometry'' in the sense that, for instance, solutions to equations satisfying the SEP, in the small body limit, follow timelike or null geodesics or ideal clocks constructed from such matter record arclength along their trajectories.    Read appears to agree when he writes,
\begin{quote}\singlespacing Why should one restrict to those solutions of GR in which the SEP is satisfied?  The reason is that this principle ... is typically regarded to constitute and important condition for the \emph{chronogeometricity} if the metric field---that is, for intervals as given by the metric field to be read off by stable rods and clocks built from matter fields. \citep[p. 10]{ReadIreny}\end{quote}
Similarly, Brown writes (and Read quotes approvingly):
\begin{quote}  \singlespacing
That light rays trace out null geodesics of the field is again a consequence of the strong equivalence principle, which asserts that locally Maxwell's equations of electrodynamics are valid. \citep[p. 176]{Brown}
\end{quote}
In other words, Read and Brown claim that a system of differential equations satisfying the SEP implies that (and explains why) (certain) solutions to that system will have certain characteristic features that we associate with relativity.

Read and Brown both assert these claims without argument.\footnote{As do others: see, for instance, \citet{KnoxFunctionalism} or \citet[pp. 18-9]{Myrvold}.}  To substantiate them, one would ideally like to formulate and prove a theorem of the form:
\begin{quote}\singlespacing Fix a system of differential equations on a Lorentzian manifold and suppose this system satisfies the SEP with respect to that metric.  Then the solutions to those equations will have the following properties: (a) in a small body limit, they will follow only timelike or null geodesics; (b) they do not propagate superluminally; (c) they have other properties characteristic of extended matter, etc. etc.\end{quote}
Of course, to prove such a theorem, one needs to do a lot of work, first and most importantly to make the SEP into a precise assertion.  But even heuristically, I am not aware of any evidence that this proposition is true.  One might reply that it, or something like it, \emph{must} be true.  I agree---but this is hardly an argument.\footnote{One might worry that insofar as my main worry about the SEP, even appropriately clarified, is the one given in this section, that ultimately my differences with Brown and Read come down to purely technical considerations, and not philosophical ones.  But I do not think this is correct.  \emph{Proving} the relevant theorems may well be a mathematical exercise, but identifying what they should be, and even stating them precisely, is a matter of deep interpretational importance with a long provenance (discussed, in part, in the final section of this paper).  I take it that the SEP reflects a particular idea about what, fundamentally, makes general relativity a theory about ``spacetime geometry'' (or, ``chronogeometry'').  As such, whether it secures the necessary results reflects on whether we truly understand what it means to attribute a certain geometry to space and time, and ultimately what the physical (and even metaphysical) significance of general relativity is.}

\section{Functionalism Revisited, or, Some Good Advice that We Just Didn't Take}

Read writes that the focus of the dynamical/geometrical debate is the following question: ``Whence the metric field's chronogeometric significance?'' \citep[p. 11]{ReadIreny}.  In reaction to this question---posed, not by Read, but by Adolf \citet{Grunbaum} some four decades before---Howard \citet{SteinGrunbaum} writes:
\begin{quote}\singlespacing The question that seems to dominate your discussion ... is this: by virtue of what is the tensor-field $g$ to be regarded as representing the metrical structure of space-time? Now that seems to me a legitimate question, but not a clear and precise one: it seems to me (to use a distinction I appear to be growing fond of) ``presystematic'' rather than ``systematic,'' and to pose a problem of explication. I should be inclined to reformulate the question as: ``Just what do we mean when we say that this tensor-field describes the metrical structure of space-time?'' \citep[p. 375]{SteinGrunbaum}\end{quote}

Stein goes on to offer his own answer to this question.
\begin{quote} \singlespacing [My answer] has a good deal to do with the theory of the behavior of such things as measuring rods and clocks. But I would begin by making  a preliminary, and in my opinion quite crucial, remark. This is that there is no ``categorical'' ... no innate, a priori, or (in your language) ``canonical'' notion of the metric of space-time. Indeed, nobody before Minkowski employed such a notion at all.   And when Minkowski invented, or discovered, this concept, what exactly did he do? He showed that the special-relativistic theory of space and time was tantamount  to the statement that space-time has a particular structure, whose attributes are suggestively (although not perfectly) analogous to those of a Euclidean metric structure ... \emph{and from which ... the Eintein geometric, chonometric, and kinematic relationships are determined}. So we have two main points: (1) a structure whose characteristics are, in a certain generalized sense, of that mathematical species which is called ``metrical''; (2) a theory according to which the physical facts that belong, classically, to physical geometry (and chronometry and kinematics) are manifestations of that structure. Geometry having thus been ``aufgehoben'' into the Minkowski structure, one quite naturally refers to the latter as ``the geometrical structure of space-time.''\citep[p. 377-8]{SteinGrunbaum}\end{quote}
I take Stein, here, to be defending precisely what I have described as the ``geometric view'' above, including endorsing what I previously referred to as (g) and (d); and to be providing an argument for why, on historical grounds, we should have understood (g) in such a way that it implies (d).

One might be tempted to despair by these passages. Has the literature on these matters made no progress in four decades?  But to the contrary, I think a great deal of progress has been made, specifically in answering Stein's preferred question: ``Just what do we mean when we say that this tensor-field describes the metrical structure of space-time?''  Stein was prepared to say that we mean that certain structures that ``measure'' geometry---small bodies, light rays, clocks---measure the properties described by the spacetime metric, and leave it at that.  But it has turned out that one can say more.  That is, we can now be more explicit, following \citet{Brown}, that what we mean by when we say that a particular tensor field describes the metrical structure of spacetime is specifically that the metric bears a particular relationship to the dynamics of these structures, or rather, the dynamics of the matter that composes them.  That relationship is manifest both in the role that the metric plays in expressions of those dynamics and in features of their solutions.  That such a relationship exists is surely in the background of Stein's remarks.  But we have made considerable progress in articulating just what it is.

So much the better for our understanding of space, time, and matter.  But, as I have argued in the forgoing, there is still much more to say about Stein's preferred question: what do we mean when we say that a particular tensor field describes the metrical structure of spacetime?  On the one hand, neither the substance nor the significance of the SEP is sufficiently clear for it to explicate the idea that matter dynamics are adapted to the spacetime metric in general relativity.  And on the other hand, although there are results in the literature that establish the sorts of links between matter field equations and our expectations about the behavior of the structures that ``measure'' spacetime geometry, those results provide only a partial picture of the complex network of relationships that one expects to hold.  Thus, we are closer to turning the ``presystematic'' questions discussed by Gr\"unbaum and Stein into systematic ones, and even ones that can be addressed by new work in mathematical physics; but we are not there yet.

Despite this progress, there is a key point in Stein's argument that has perhaps been overlooked in the subsequent literature, but which I think is reflected in some of the difficulties I have just highlighted.  It is a problem not so much for the dynamic view, but rather for the ``extension'' to that view noted in the introduction to this paper: spacetime functionalisnm.\footnote{As noted above, \citet[fn. 40]{WeatherallConservation} offers some considerations for why one should not be an inertial frame functionalist; \citet{Read+Menon} offer other arguments for why inertial frame functionalism has limitations.  The concerns I raise here are not about inertial frame functionalism in particular, but rather the idea that one can say in advance what we even mean by ``spacetime role'' once and for all; identifying that role with inertial frames is just one example.  \citet{Baker} argues that the spacetime concept is a ``clus ter concept'', in the sense that there are many different, related conditions that one would associate with it without any of them being jointly necessary or sufficient for it to apply.  That argument is closer to the one I raise here, except that my emphasis is on how difficult and theory-dependent it is to identify what we mean by ``the spacetime role''.}  Of course, one worry for this view, at least as defended by \citet{KnoxFunctionalism}, that immediately rises from the forgoing discussion, is that Knox also invokes the SEP as her justification that it is the metric in general relativity that \emph{does} play the spacetime role; for the reasons already noted, I think this is a problematic claim.  But that is not my concern.  Rather, what I wish to emphasize is that even within general relativity, a theory we have been working with for a century, to clearly and precisely articulate what it means for some structure to ``play the spacetime role'' or to have ``chronogeometric significance'' is subtle and difficult, and it is not clear that we have control over it.  As Stein puts it, ``there is no ... `canonical' notion of the metric of space-time.''  In other words, the expectation that a particular field, $g_{ab}$, \emph{does} play a certain ``spatio-temporal'' or ``chronogeometric'' role, or has a certain significance, arose only within the context of developing this particular theory.  The very idea of spacetime having a geometry in the relevant sense---of their being a ``spacetime role'' to play here at all---did not make sense before special relativity came along, and changed dramatically once general relativity was introduced.

Looking backward, one can ask various questions along the lines of the ones discussed here, and identify conditions, within general relativity and closely related theories, that would be sufficient for a structure to ``play the spacetime role''.  But the history of how these concepts came about, and the difficulties associated with making them clear, make me pessimistic that we can now say, once and for all, what it means to ``play the spacetime role'' in a theory-independent way.  Yet this is precisely what functionalism seeks to do: it is an account of what spacetime is, and how we get to identify a structure as spacetime given the dynamics of matter, independent of any theory.\footnote{Some uses of functionalism in the recent literature, such as that of \citet{KnoxEffective} and \citet{Wuthrich+Lam}, have addressed a somewhat different question, namely, what would need to be the case for spacetime, in the sense of general relativity, to ``emerge'' from a different theory, such as a theory with different geometrical structure (Knox) or a quantum theory of gravity (W\"uthrich and Lam).  In this case, since general relativity is the target, the concern that I raise here does not get purchase.}  Indeed, as evidence for why one ought to be pessimistic about this program, just observe how difficult it is to go from considerations that make sense in special relativity, concerning, say, the Poincar\'e invariance of dynamical equations (in the sense of solutions of equations being mapped to other solutions by automorphisms of Minkowski spacetime), to analogous considerations in general relativity.  Why not expect that to articulate the ``spacetime role'' in a successful theory of quantum gravity will require even more complexity, or else draw on completely different concepts that we will need to develop in the course of developing that theory?

Here, I think, we see an important distinction between two different claims with a functionalist flavor.  One is the claim that it is because the metric and its associated derivative operator play certain roles in the dynamics of matter that that matter has the dynamical properties (say, in the small-body limit) that one expects in general relativity.  This claim, which amounts to a line on what I called (B) above, could be substantiated in greater detail, but seems almost certain to be true.  But the second claim is that it is by virtue of playing a particular, pre-specified role in a theory, irrespective of what that theory is, that a structure counts as spacetime.  This is clearly a much stronger claim, and one that I think should be approached with caution.

\section*{Acknowledgments}
This material is partially based upon work produced for the project ``New Directions in Philosophy of Cosmology'', funded by the John Templeton Foundation under grant number 61048. I am grateful to Sam Fletcher, David Malament, and James Read for comments on an earlier draft of this manuscript, to Chris W\"uthrich for encouraging me to produce it, to two anonymous referees for helpful suggestions, and to Clara Bradley, Harvey Brown, John Earman, Sam Fletcher, Eleanor Knox, Dennis Lehmkuhl, David Malament, Oliver Pooley, and David Wallace for many conversations over the years related to this material.

\singlespacing

\bibliographystyle{elsarticle-harv}
\bibliography{geometric}

\begin{thebibliography}{48}
\expandafter\ifx\csname natexlab\endcsname\relax\def\natexlab#1{#1}\fi
\expandafter\ifx\csname url\endcsname\relax
  \def\url#1{\texttt{#1}}\fi
\expandafter\ifx\csname urlprefix\endcsname\relax\def\urlprefix{URL }\fi

\bibitem[{Acu{\~n}a(2016)}]{Acuna}
Acu{\~n}a, P., 2016. Minkowski spacetime and lorentz invariance: The cart and
  the horse or two sides of a single coin? Studies in History and Philosophy of
  Science Part B: Studies in History and Philosophy of Modern Physics 55,
  1--12.

\bibitem[{Baker(2019)}]{Baker}
Baker, D.~J., 2019. On spacetime functionalism, . Unpublished manuscript.
  http://philsci-archive.pitt.edu/15860/.

\bibitem[{Brown(2005)}]{Brown}
Brown, H.~R., 2005. Physical Relativity. Oxford University Press, New York.

\bibitem[{Brown and Pooley(1999)}]{Brown+Pooley1}
Brown, H.~R., Pooley, O., 1999. The origins of the spacetime metric: Bell's
  lorentzian pedagogy and its significance in general relativity. In:
  Callender, C., Huggett, N. (Eds.), Physics Meets Philosophy at the Planck
  Scale. Cambridge University Press, Cambridge, UK, pp. 256--72.

\bibitem[{Brown and Pooley(2006)}]{Brown+Pooley2}
Brown, H.~R., Pooley, O., 2006. Minkowski space-time: A glorious non-entity.
  In: Dieks, D. (Ed.), The Ontology of Spacetime. Elsevier, Amsterdam, pp.
  67--89.

\bibitem[{Brown and Read(20??)}]{Brown+Read}
Brown, H.~R., Read, J., 20?? The dynamical approach to spacetime theories. In:
  Knox, E., Wilson, A. (Eds.), The Routledge Companion to Philosophy of
  Physics. Routledge, London, . Forthcoming.
  http://philsci-archive.pitt.edu/14592/.

\bibitem[{Brown and Read(2016)}]{Misconceptions}
Brown, H.~R., Read, J., 2016. Clarifying possible misconceptions in the
  foundations of general relativity. American Journal of Physics 84~(5),
  327--334.

\bibitem[{Curiel(2017)}]{CurielEC}
Curiel, E., 2017. A primer on energy conditions. In: Lehmkuhl, D., Schiemann,
  G., Scholz, E. (Eds.), Towards a Theory of Spacetime Theories. Birkh\"auser,
  Boston, MA, pp. 43--104.

\bibitem[{Dewar(2020)}]{Dewar}
Dewar, N., 2020. General-relativistic covariance. Foundations of Physics 50,
  294--318.

\bibitem[{Dorato(2007)}]{Dorato}
Dorato, M., 2007. Relativity theory between structural and dynamical
  explanations. International Studies in the Philosophy of Science 21~(1),
  95--102.

\bibitem[{Earman(2014)}]{EarmanSL}
Earman, J., 2014. No superluminal propagation for classical relativistic and
  relativistic quantum fields. Studies in History and Philosophy of Science
  Part B: Studies in History and Philosophy of Modern Physics 48, 102--108.

\bibitem[{Fletcher(20??)}]{Fletcher}
Fletcher, S.~C., 20?? Approximate local poincar\'e spacetime symmetry in
  general relativity. In: Beisbart, C., Sauer, T., W\"uthrich, C. (Eds.),
  Thinking About Space and Time: 100 Years of Applying and Interpreting General
  Relativity. Birkh\"auser, Boston, MA, . Forthcoming.

\bibitem[{Fletcher(2013)}]{FletcherCH}
Fletcher, S.~C., 2013. Light clocks and the clock hypothesis. Foundations of
  Physics 43~(11), 1369--1383.

\bibitem[{Fletcher and Weatherall(2020)}]{Fletcher+Weatherall}
Fletcher, S.~C., Weatherall, J.~O., 2020. Is spacetime approximately locally
  flat?, unpublished manuscript.

\bibitem[{Friedman(1983)}]{Friedman}
Friedman, M., 1983. Foundations of Space-Time Theories: Relativistic Physics
  and Philosophy of Science. Princeton University Press, Princeton, NJ.

\bibitem[{Frisch(2011)}]{Frisch}
Frisch, M., 2011. Principle or constructive relativity. Studies in History and
  Philosophy of Modern Physics 42~(3), 176--183.

\bibitem[{Geroch(2011)}]{GerochFTL}
Geroch, R., 2011. Faster than light? In: Plaue, M., Rendall, A., Scherfner, M.
  (Eds.), Advances in Lorentzian Geometry. American Mathematical Society,
  Providence, RI, pp. 59--80.

\bibitem[{Geroch and Weatherall(2018)}]{Geroch+Weatherall}
Geroch, R., Weatherall, J.~O., 2018. The motion of small bodies in space-time.
  Communications in Mathematical Physics 364, 607--634,
  dOI:10.1007/s00220-018-3268-8.

\bibitem[{Gr\"unbaum(1977)}]{Grunbaum}
Gr\"unbaum, A., 1977. Absolute and relational theories of space and space-time.
  In: Earman, J., Glymour, C., Stachel, J. (Eds.), Foundations of Space-Time
  Theories. University of Minnesota Press, Minneapolis, MN, pp. 303--373.

\bibitem[{Hawking and Ellis(1973)}]{Hawking+Ellis}
Hawking, S.~W., Ellis, G. F.~R., 1973. The Large Scale Structure of Space-time.
  Cambridge University Press, New York.

\bibitem[{Janssen(2009)}]{Janssen}
Janssen, M., 2009. Drawing the line between kinematics and dynamics in special
  relativity. Studies in History and Philosophy of Science Part B: Studies in
  History and Philosophy of Modern Physics 40~(1), 26--52.

\bibitem[{Knox(2013)}]{KnoxEffective}
Knox, E., 2013. Effective spacetime geometry. Studies in History and Philosophy
  of Science Part B: Studies in History and Philosophy of Modern Physics
  44~(3), 346--356.

\bibitem[{Knox(2019)}]{KnoxFunctionalism}
Knox, E., 2019. Physical relativity from a functionalist perspective. Studies
  in History and Philosophy of Modern Physics 67, 118--124.

\bibitem[{Lam and W\"uthrich(2018)}]{Wuthrich+Lam}
Lam, V., W\"uthrich, C., 2018. Spacetime is as spacetime does. Studies in
  History and Philosophy of Modern Physics 64, 39--51.

\bibitem[{Malament(2012{\natexlab{a}})}]{MalamentGP}
Malament, D., 2012{\natexlab{a}}. A remark about the ``geodesic principle'' in
  general relativity. In: Frappier, M., Brown, D.~H., DiSalle, R. (Eds.),
  Analysis and Interpretation in the Exact Sciences: Essays in Honour of
  William Demopoulos. Springer, New York, pp. 245--252.

\bibitem[{Malament(2012{\natexlab{b}})}]{MalamentGR}
Malament, D.~B., 2012{\natexlab{b}}. Topics in the Foundations of General
  Relativity and Newtonian Gravitation Theory. University of Chicago Press,
  Chicago, IL.

\bibitem[{Maudlin(2012)}]{Maudlin}
Maudlin, T., 2012. Philosophy of Physics: Space and Time. Princeton University
  Press, Princeton, NJ.

\bibitem[{Misner et~al.(1973)Misner, Thorne, and Wheeler}]{MTW}
Misner, C.~W., Thorne, K.~S., Wheeler, J.~A., 1973. Gravitation. W. H. Freeman.

\bibitem[{Myrvold(2019)}]{Myrvold}
Myrvold, W.~C., 2019. How could relativity be anything other than physical?
  Studies in History and Philosophy of Science Part B: Studies in History and
  Philosophy of Modern Physics 67, 137--143.

\bibitem[{Norton(2008)}]{Norton}
Norton, J.~D., 2008. Why constructive relativity fails. The British Journal for
  the Philosophy of Science 59, 821--834.

\bibitem[{Read(20??)}]{ReadIreny}
Read, J., 20?? Explanation, geometry, and conspiracy in relativity theory. In:
  Beisbart, C., Sauer, T., W\"uthrich, C. (Eds.), Thinking About Space and
  Time: 100 Years of Applying and Interpreting General Relativity.
  Birkh\"auser, Boston, MA, . Forthcoming.

\bibitem[{Read(2019)}]{ReadConstructive}
Read, J., 2019. Geomertrical constructivism and modal relationism: Further
  aspects of the dynamical/geometrical debate, unpublished ms.

\bibitem[{Read et~al.(2018)Read, Brown, and Lehmkuhl}]{Miracles}
Read, J., Brown, H.~R., Lehmkuhl, D., 2018. Two miracles of general relativity.
  Studies in History and Philosophy of Science Part B: Studies in History and
  Philosophy of Modern Physics 64, 14--25.

\bibitem[{Read and Menon(2019)}]{Read+Menon}
Read, J., Menon, T., 2019. The limitations of inertial frame spacetime
  functionalism. Synthese. DOI: 10.1007/s11229-019-02299-2.

\bibitem[{Stein(1977)}]{SteinGrunbaum}
Stein, H., 1977. On space-time and ontology: Extract from a letter to adolf
  gr\"unbaum. In: Earman, J., Glymour, C., Stachel, J. (Eds.), Foundations of
  Space-Time Theories. University of Minnesota Press, Minneapolis, MN, pp.
  374--402.

\bibitem[{Sus(2014)}]{Sus}
Sus, A., 2014. On the explanation of inertia. Journal for General Philosophy of
  Science 45~(2), 293--315.

\bibitem[{Torretti(1983)}]{Torretti}
Torretti, R., 1983. Relativity and Geometry. Pergamon Press, Oxford, UK.

\bibitem[{Wald(1984)}]{Wald}
Wald, R.~M., 1984. General Relativity. University of Chicago Press, Chicago.

\bibitem[{Wallace(2019)}]{WallaceWACS}
Wallace, D., 2019. Who's afraid of coordinate systems? an essay on
  representation of spacetime structure. Studies in history and philosophy of
  modern physics 67, 125--136.

\bibitem[{Weatherall(20??)}]{WeatherallGM}
Weatherall, J.~O., 20?? Geometry and motion in general relativity. In:
  Beisbart, C., Sauer, T., W\"uthrich, C. (Eds.), Thinking About Space and
  Time: 100 Years of Applying and Interpreting General Relativity.
  Birkh\"auser, Boston, MA, . Forthcoming.

\bibitem[{Weatherall(2011)}]{WeatherallGP}
Weatherall, J.~O., 2011. On the status of the geodesic principle in {N}ewtonian
  and relativistic physics. Studies in the History and Philosophy of Modern
  Physics 42~(4), 276--281.

\bibitem[{Weatherall(2014)}]{WeatherallSL}
Weatherall, J.~O., 2014. Against dogma: On superluminal propagation in
  classical electromagnetism. Studies in History and Philosophy of Science Part
  B: Studies in History and Philosophy of Modern Physics 48, 109--123.

\bibitem[{Weatherall(2016)}]{WeatherallHoleArg}
Weatherall, J.~O., 2016. Regarding the ‘hole argument’. The British Journal
  for the Philosophy of Science 69~(2), 329--350.

\bibitem[{Weatherall(2017)}]{WeatherallPuzzleball}
Weatherall, J.~O., 2017. Inertial motion, explanation, and the foundations of
  classical space-time theories. In: Lehmkuhl, D., Schiemann, G., Scholz, E.
  (Eds.), Towards a Theory of Spacetime Theories. Birkh\"auser, Boston, MA, pp.
  13--42.

\bibitem[{Weatherall(2019)}]{WeatherallConservation}
Weatherall, J.~O., 2019. Conservation, inertia, and spacetime geometry. Studies
  in History and Philosophy of Science Part B: Studies in History and
  Philosophy of Modern Physics 67, 144--159.

\bibitem[{Weatherall(2020)}]{WeatherallStein}
Weatherall, J.~O., 2020. Some philosophical prehistory of the
  ({E}arman-{N}orton) hole argument. Studies in History and Philosophy of
  Modern Physics 70, 79--87.

\bibitem[{Weyl(1952 [1918])}]{Weyl}
Weyl, H., 1952 [1918]. Space Time Matter. Dover Publications, Mineola, NY.

\bibitem[{Wilson(2006)}]{Wilson}
Wilson, M., 2006. Wandering Significance: An Essay on Conceptual Behavior.
  Oxford University Press, Oxford, UK.

\end{thebibliography}

\end{document}